\documentclass[11pt]{article}
\pdfoutput=1
\usepackage{amsmath,amsfonts,amssymb}
\usepackage{cite,url}
\usepackage{graphicx}
\usepackage[utf8]{inputenc} 
\usepackage[bottom]{footmisc}

\allowdisplaybreaks

\def\smallZ{{\scriptscriptstyle Z}}
\def\smallW{{\scriptscriptstyle W}}

\def\smallL{{\scriptscriptstyle L}}
\def\smallR{{\scriptscriptstyle R}}

\def\smallEW{{\scriptscriptstyle EW}}
\def\xqu{x_{\scriptscriptstyle QU}}
\def\MS{M_S}
\def\MZ{m_\smallZ}
\def\MW{m_\smallW}

\def\beq{\begin{equation}}
\def\eeq{\end{equation}}
\def\bea{\begin{eqnarray}}
\def\eea{\end{eqnarray}}
\def\nn{\nonumber}
\def\wt{\widetilde}

\def\dWFR{\Delta_{\rm {\scriptscriptstyle WFR}}}
\def\qcdew{QCD--EW}

\def\cdbe{c_{2\beta}}

\def\sq2{\sqrt{2}}
\def\drbar{{\ensuremath{ \overline{\rm DR}}}}
\def\msbar{\overline{\rm MS}}
\def\smalldrbar{\scriptscriptstyle{\overline{\rm DR}}}
\def\smallmsbar{\scriptscriptstyle{\overline{\rm MS}}}

\def\smalldred{{\scriptscriptstyle {\rm DRED}}}
\def\smalldreg{{\scriptscriptstyle {\rm DREG}}}

\def\tb{\tan\beta}
\def\gl{\tilde{g}}
\def\mg{m_{\gl}}
\def\g{\mg^2}
\def\mh{m_h}

\def\qew{Q_{{\rm{\scriptscriptstyle EW}}}}

\def\hgz{\hat{g}_\smallZ}
\def\gz{g_\smallZ}
\def\gs{g_s}

\def\gp{g^\prime}
\def\gpq{g^{\prime\,2}}
\def\gpqq{g^{\prime\,4}}

\def\mt{m_t}

\def\t{\mt^2}

\def\tu{m_{\tilde{t}_1}^2}
\def\td{m_{\tilde{t}_2}^2}
\def\tul{m_{\tilde{t}_1}}
\def\tdl{m_{\tilde{t}_2}}

\def\sdt{s_{2\theta_t}}
\def\cdt{c_{2\theta_t}}

\def\mb{m_b}

\def\msqq{m_{\tilde q}^2}

\long\def\symbolfootnote[#1]#2{\begingroup%
\def\thefootnote{\fnsymbol{footnote}}\footnote[#1]{#2}\endgroup}

\makeatletter
\newcommand{\vast}{\bBigg@{3}}
\makeatother

\begin{document}

\begin{titlepage}

\begin{flushright}
{\tt PSI-PR-19-14}
\end{flushright}

\vspace{1cm}
\begin{center}

\vspace{1cm}

{\LARGE \bf Full two-loop QCD corrections to the Higgs mass} 
\vskip 0.3cm
{\LARGE \bf in the MSSM with heavy superpartners}

\vspace{1cm}

{\Large Emanuele Bagnaschi,$^{\!\!\!\,a}$~ 
  Giuseppe Degrassi,$^{\!\!\!\,b,\,c}$ ~Sebastian Pa{\ss}ehr$^{\,d}$}

\vspace*{2mm}

{\Large   and Pietro~Slavich$^{\,d}$}

\vspace*{5mm}

{\sl ${}^a$ Paul Scherrer Institut, CH-5232 Villigen PSI, Switzerland}
\vspace*{2mm}\\{\sl ${}^b$ 
Dipartimento di Matematica e Fisica, Università di Roma Tre

Via della Vasca Navale 84, I-00146 Rome, Italy}
\vspace*{2mm}\\{\sl ${}^c$ INFN, Sezione di Roma Tre, Via della Vasca Navale 84, I-00146 Rome, Italy}
\vspace*{2mm}\\{\sl ${}^d$
  Laboratoire de Physique Th\'eorique et Hautes Energies (LPTHE),
  
  UMR 7589, Sorbonne Université et CNRS, 4 place Jussieu,
  75252 Paris Cedex 05, France.}
\end{center}
\symbolfootnote[0]{{\tt e-mail:}}
\symbolfootnote[0]{{\tt emanuele.bagnaschi@psi.ch}}
\symbolfootnote[0]{{\tt degrassi@fis.uniroma3.it}}
\symbolfootnote[0]{{\tt passehr@lpthe.jussieu.fr}}
\symbolfootnote[0]{{\tt slavich@lpthe.jussieu.fr}}

\vspace{0.7cm}

\abstract{We improve the determination of the Higgs-boson mass in the
  MSSM with heavy superpartners, by computing the two-loop threshold
  corrections to the quartic Higgs coupling that involve both the
  strong and the electroweak gauge couplings. Combined with earlier
  results, this completes the calculation of the two-loop QCD
  corrections to the quartic coupling at the SUSY scale.  We also
  compare different computations of the relation between the quartic
  coupling and the pole mass of the Higgs boson at the EW scale. We
  find that the numerical impact of the new corrections on the
  prediction for the Higgs mass is modest, but comparable to the
  accuracy of the Higgs-mass measurement at the LHC.}

\vfill

\end{titlepage}


\setcounter{footnote}{0}

\section{Introduction}
\label{sec:intro}
The Minimal Supersymmetric Standard Model (MSSM) is one of the
best-motivated extensions of the Standard Model (SM), and probably the
most studied. The Higgs sector of the MSSM consists of two $SU(2)$
doublets, but the model allows for a so-called ``decoupling limit'' in
which a combination of the two doublets has SM-like couplings to
matter fermions and gauge bosons -- so that its neutral scalar
component $h$ can be identified with the Higgs boson discovered at the
LHC~\cite{Aad:2012tfa, Chatrchyan:2012xdj}, which itself is broadly
SM-like~\cite{Khachatryan:2016vau} -- while the orthogonal combination
of doublets is much heavier. An important aspect of the MSSM is the
existence of relations between the quartic Higgs couplings and the
electroweak (EW) gauge couplings.  In the decoupling limit, these
relations induce a tree-level prediction $(\mh^2)^{\rm tree} \approx
\MZ^2\cos^2 2\beta$ for the squared mass of the SM-like scalar, where
$\MZ$ is the $Z$-boson mass, and the angle~$\beta$ is related to the
ratio of the vacuum expectation values (vevs) of the two Higgs
doublets by $\tan\beta =$~$v_2/v_1$, and determines their admixture into
$h$. Consequently, in the MSSM, the tree-level contribution can only
make up for at most half of the squared mass of the observed Higgs
boson, $(\mh^2)^{\rm obs} \approx (125~{\rm
  GeV})^2$~\cite{Aad:2015zhl}. The rest must arise from radiative
corrections. It has been known since the early
1990s~\cite{Okada:1990vk, Ellis:1990nz, Haber:1990aw, Okada:1990gg,
  Ellis:1991zd, Brignole:1991pq} that the most relevant corrections to
the Higgs mass are those controlled by the top Yukawa coupling, $g_t
\sim {\cal O}(1)$, which involve the top quark and its superpartners,
the stop squarks. These corrections are enhanced by logarithms of the
ratio between stop and top masses, and also show a significant
dependence on the value of the left--right stop mixing parameter
$X_t$. In particular, for values of $\tan\beta$ large enough to
saturate the tree-level prediction, a Higgs mass around $125$~GeV can
be obtained with an average stop mass $\MS$ of about $1$--$2$~TeV when
$X_t/\MS \approx 2$, whereas for vanishing~$X_t$ the stops need to be
heavier than $10$~TeV.

Over the years, the crucial role of the radiative corrections
stimulated a wide effort to compute them with the highest possible
precision, in order to keep the theoretical uncertainty of the
Higgs-mass prediction under control. By now, that computation is
indeed quite advanced: full one-loop
corrections~\cite{Chankowski:1991md, Brignole:1991wp, Brignole:1992uf,
  Chankowski:1992er, Dabelstein:1994hb, Pierce:1996zz} and two-loop
corrections in the limit of vanishing external
momentum~\cite{Hempfling:1993qq, Heinemeyer:1998jw, Heinemeyer:1998kz,
  Zhang:1998bm, Heinemeyer:1998np, Espinosa:1999zm, Espinosa:2000df,
  Degrassi:2001yf, Brignole:2001jy, Brignole:2002bz, Martin:2002iu,
  Martin:2002wn, Dedes:2003km, Heinemeyer:2004xw} are available, and
the dominant momentum-dependent two-loop
corrections~\cite{Martin:2003qz, Martin:2003it, Martin:2004kr,
  Borowka:2014wla, Degrassi:2014pfa} as well as the dominant
three-loop corrections~\cite{Harlander:2008ju, Kant:2010tf,
  Harlander:2017kuc, Stockinger:2018oxe, R.:2019ply, R.:2019irs} have
also been obtained.\footnote{\,We focus here on the MSSM with real
  parameters. Significant efforts have also been devoted to the
  Higgs-mass calculation in the presence of CP-violating
  phases~\cite{Pilaftsis:1999qt, Choi:2000wz, Carena:2000yi,
    Frank:2006yh, Heinemeyer:2007aq, Hollik:2014wea, Hollik:2014bua,
    Hollik:2015ema, Goodsell:2016udb, Passehr:2017ufr,
    Borowka:2018anu}, as well as in non-minimal SUSY extensions of the
  SM.} However, when the SUSY scale~$\MS$ is significantly larger than
the EW scale (which we can identify, e.g., with the top mass $\mt$),
any fixed-order computation of $\mh$ may become inadequate, because
radiative corrections of order $n$ in the loop expansion contain terms
enhanced by as much as $\ln^n(\MS/\mt)$. In the presence of a
significant hierarchy between the scales, the computation of the Higgs
mass needs to be reorganized in an effective field theory (EFT)
approach: the heavy particles are integrated out at the scale $\MS$,
where they only affect the matching conditions for the couplings of
the EFT valid below $\MS$; the appropriate renormalization group
equations (RGEs) are then used to evolve those couplings between the
SUSY scale and the EW scale, where the running couplings are related
to physical observables such as the Higgs-boson mass and the masses of
fermions and gauge bosons.  In this approach, the computation is free
of large logarithmic terms both at the SUSY scale and at the EW scale,
while the effect of those terms is accounted for to all orders in the
loop expansion by the evolution of the couplings between the two
scales. More precisely, large corrections can be resummed to the
(next-to)$^n$-leading-logarithmic (N$^n$LL) order by means of $n$-loop
calculations at the SUSY and EW scales combined with $(n\!+\!1)$-loop
RGEs.

The EFT approach to the computation of the MSSM Higgs mass dates back
to the early 1990s~\cite{Barbieri:1990ja, Espinosa:1991fc,
  Casas:1994us}, and it has also been exploited in the
past~\cite{Haber:1993an,Carena:1995bx, Carena:1995wu, Haber:1996fp,
  Carena:2000dp, Degrassi:2002fi, Martin:2007pg} to determine
analytically the coefficients of the logarithmic terms in the
Higgs-mass corrections, by solving perturbatively the appropriate
systems of boundary conditions and RGEs. In recent years, after the
LHC results pushed the expectations for the SUSY scale into the TeV
range, the realization that an accurate prediction for the Higgs mass
in the MSSM cannot prescind from the resummation of the large
logarithmic corrections brought the EFT computation under renewed
focus~\cite{Hahn:2013ria, Draper:2013oza, Bagnaschi:2014rsa,
  Vega:2015fna, Lee:2015uza, Bagnaschi:2015pwa, Bahl:2016brp,
  Athron:2016fuq, Staub:2017jnp, Bagnaschi:2017xid, Bahl:2017aev,
  Athron:2017fvs, Allanach:2018fif, Bahl:2018jom, Harlander:2018yhj,
  Bahl:2018ykj}. In the simplest scenario in which all of the SUSY
particles as well as the heavy Higgs doublet of the MSSM are clustered
around a single scale $\MS$, so that the EFT valid below that scale is
just the SM, the state of the art now includes: full one-loop and
partial two-loop matching conditions for the quartic Higgs coupling at
the SUSY scale, computed for arbitrary values of the relevant SUSY
parameters~\cite{Bagnaschi:2014rsa, Bagnaschi:2017xid}; full
three-loop RGEs for all of the parameters of the SM
Lagrangian~\cite{Mihaila:2012fm, Chetyrkin:2012rz, Mihaila:2012pz,
  Bednyakov:2012en, Chetyrkin:2013wya, Bednyakov:2013eba}; full
two-loop relations at the EW scale between the running SM parameters
and a set of physical observables which include the pole Higgs
mass~\cite{Buttazzo:2013uya, Kniehl:2015nwa, Martin:2019lqd}. The
combination of these results allows for a full NLL resummation of the
large logarithmic corrections to the Higgs mass, whereas the NNLL
resummation can only be considered partial, because in
refs.~\cite{Bagnaschi:2014rsa, Bagnaschi:2017xid} the two-loop
matching conditions for the quartic Higgs coupling were computed in
the ``gaugeless limit'' of vanishing EW gauge couplings.\footnote{A
  partial N$^3$LL resummation of the corrections involving only the
  highest powers of the strong gauge coupling is also available,
  combining the three-loop matching condition of
  ref.~\cite{Harlander:2018yhj} with SM results from
  refs.~\cite{Bednyakov:2013eba, Martin:2013gka, Martin:2014cxa,
    Martin:2015eia}.}

Beyond the pure EFT calculation of the Higgs mass, different
``hybrid'' approaches to combine the existing ``diagrammatic'' (i.e.,
fixed-order) calculations with a resummation of the logarithmic
corrections have been proposed~\cite{Hahn:2013ria, Bahl:2016brp,
  Athron:2016fuq, Staub:2017jnp}. The aim is to include terms
suppressed by powers of $v^2/\MS^2\,$ (where we denote by $v$ the vev
of a SM-like Higgs scalar) up to the perturbative order accounted for
by the diagrammatic calculation. In the EFT calculation, those terms
can be mapped to the effect of non-renormalizable, higher-dimensional
operators, and they are neglected when the theory valid below the
matching scale is taken to be the plain SM in the unbroken phase of
the EW symmetry. To avoid double counting, the hybrid approaches
require a careful subtraction of the terms that are accounted for by
both the diagrammatic and the EFT calculations, and indeed a few
successive adjustments~\cite{Bahl:2017aev, Athron:2017fvs,
  Bahl:2018ykj} were necessary to obtain predictions for $\mh$ that,
in the limit of very heavy SUSY masses in which the ${\cal
  O}(v^2/\MS^2)$ terms are certainly negligible, show the expected
agreement with the pure EFT calculation. The comparison between the
predictions of the hybrid and pure EFT calculations, as well as a
direct study~\cite{Bagnaschi:2017xid} of the effects of
non-renormalizable operators in the EFT, also show that the ${\cal
  O}(v^2/\MS^2)$ corrections are significantly suppressed for the
values of $\MS$ that are large enough to allow for $\mh \approx
125$~GeV. Other recent developments of the EFT approach include: the
study of MSSM scenarios in which both Higgs doublets are light, so
that the effective theory valid below the SUSY scale is a
two-Higgs-doublet model (THDM)~\cite{Lee:2015uza, Bagnaschi:2015pwa,
  Bahl:2018jom}; the application of functional techniques to the full
one-loop matching of the MSSM onto the SM~\cite{Wells:2017vla}; the
calculation of one-loop matching conditions between the couplings of
two generic renormalizable theories~\cite{Braathen:2018htl,
  Gabelmann:2018axh}, which can then be adapted to SUSY (or non-SUSY)
models other than the MSSM.

In this paper we focus again on the simplest EFT setup in which the
theory valid below the SUSY scale is the SM, and we take a further
step towards the full NNLL resummation of the large logarithmic
corrections. In particular, we compute the two-loop threshold
corrections to the quartic Higgs coupling that involve both the strong
and the EW gauge couplings. Combined with the ``gaugeless'' results of
refs.~\cite{Bagnaschi:2014rsa, Bagnaschi:2017xid}, this completes the
calculation of the two-loop threshold corrections that involve the
strong gauge coupling. We also discuss the necessary inclusion of
contributions beyond the gaugeless limit in the relation between the
pole Higgs mass and the $\msbar$-renormalized quartic Higgs coupling
at the EW scale, and we compare the results of the full two-loop
calculations of that relation given in refs.~\cite{Buttazzo:2013uya}
and \cite{Kniehl:2015nwa}, respectively.  Finally, considering a
representative scenario for the MSSM with heavy superpartners, we find
that the numerical impact of the new corrections on the prediction for
the Higgs mass is modest, but comparable to the accuracy of the
Higgs-mass measurement at the LHC.

\vspace*{-1.5mm}
\section{Two-loop matching of the quartic Higgs coupling}
\label{sec:matching}

In this section we describe our calculation of the two-loop QCD
contributions to the matching condition for the quartic Higgs
coupling. We consider the setup in which all SUSY particles as well as
a linear combination of the two Higgs doublets of the MSSM are
integrated out at a common renormalization scale $Q\approx\MS$, so
that the EFT valid below the matching scale is the SM. In our
conventions the potential for the SM-like Higgs doublet $H$ contains
the quartic interaction term $\frac\lambda 2 \,|H|^4$, and the
tree-level squared mass of its neutral scalar component is
$(\mh^2)^{\rm tree} = 2\lambda v^2$, with $v = \langle H^0 \rangle
\approx 174$~GeV. Then the two-loop matching condition for the quartic
coupling takes the form
\beq
\label{looplam}
\lambda (Q)= \frac14\left[g^2(Q)+ g^{\prime\,2}(Q)\right] 
\cdbe^2 (Q)
~+~ \Delta \lambda^{1\ell}
~+~ \Delta \lambda^{2\ell}~,  
\eeq
where $g$ and $\gp$ are the EW gauge couplings, $\beta$ can be
interpreted as the angle\,\footnote{Here and thereafter, we use the
  shortcuts $c_\phi\equiv\cos\phi$ and $s_\phi\equiv\sin\phi$ for a
  generic angle $\phi$.} that rotates the two original MSSM doublets
into a light doublet $H$ and a massive doublet $A$, and $\Delta
\lambda^{n\ell}$ is the $n$-loop threshold correction to the quartic
coupling arising from integrating out the heavy particles at the scale
$Q$. The complete result for the one-loop correction $\Delta
\lambda^{1\ell}$, valid for arbitrary values of all the relevant SUSY
parameters, can be found in refs.~\cite{Bagnaschi:2014rsa,
  Bagnaschi:2017xid}. It is computed under the assumptions that
$\lambda$, $g$ and $\gp$ in eq.~(\ref{looplam}) are
$\msbar$-renormalized parameters of the SM, and that $\beta$ is
defined beyond tree level as described in section 2.2 of
ref.~\cite{Bagnaschi:2014rsa}, removing entirely the contributions of
the off-diagonal wave-function renormalization (WFR) of the Higgs
doublets. As to the two-loop correction $\Delta \lambda^{2\ell}$,
ref.~\cite{Bagnaschi:2014rsa} provided the contributions of ${\cal
  O}(g_t^4\,\gs^2)$, where $\gs$ is the strong gauge coupling, for
arbitrary values of all the relevant SUSY parameters;
ref.~\cite{Bagnaschi:2017xid} provided in addition the two-loop
contributions involving only the third-family Yukawa couplings, $g_t$,
$g_b$ and $g_\tau$, again for arbitrary SUSY parameters, and also
discussed some subtleties in the derivation of the ${\cal
  O}(g_b^4\,\gs^2)$ contributions from the known results for the
${\cal O}(g_t^4\,\gs^2)$ ones. Altogether, the results of
refs.~\cite{Bagnaschi:2014rsa, Bagnaschi:2017xid} amounted to a
complete determination of $\Delta \lambda^{2\ell}$ in the limit of
vanishing EW gauge (and first-two-generation Yukawa) couplings. In
this paper we take a step beyond this ``gaugeless limit'' and compute
the remaining corrections that involve the strong gauge coupling,
namely those of ${\cal O}(g_{t,b}^2\,g^2 \gs^2)$, those of ${\cal
  O}(g_{t,b}^2 \, \gpq \gs^2)$ and those involving only gauge
couplings, i.e.~of ${\cal O}(g^4 \gs^2)$ and ${\cal O}(\gpqq \gs^2)$.

\bigskip

We can decompose the ``mixed'' \qcdew\  threshold correction to the
quartic Higgs coupling into three terms:
\beq
\label{threeparts}
\Delta \lambda^{2\ell,\,{\rm {\scriptscriptstyle QCD\mbox{-}EW}}} ~=~
\Delta \lambda^{2\ell,{\rm 1{\scriptscriptstyle PI}}}
\,+~  \Delta \lambda^{2\ell,{\rm {\scriptscriptstyle WFR}}}
\,+~  \Delta \lambda^{2\ell,\,{\rm {\scriptscriptstyle RS}}}~.
\eeq
The first term on the r.h.s.~of the equation above denotes the
contributions of two-loop, one-particle-irreducible (1PI) diagrams
with four external Higgs fields involving both strong-interaction
vertices (namely squark--gluon, four-squark or quark--squark--gluino
vertices) and $D$-term-induced quartic Higgs--squark vertices
proportional to $g^2$ or to $\gpq$, and possibly also vertices
controlled by the Yukawa couplings. This term can be computed with a
relatively straightforward adaptation of the effective-potential
approach developed in refs.~\cite{Bagnaschi:2014rsa,
  Bagnaschi:2017xid} for the corresponding calculation in the
gaugeless limit. In contrast, the remaining terms on the r.h.s.~of
eq.~(\ref{threeparts}) require different approaches.
The second term involves the two-loop, ${\cal O}(g_{t,b}^2 \, \gs^2)$
squark contributions to the WFR of the Higgs field, which multiply the
tree-level quartic coupling, see eq.~(\ref{looplam}), giving rise to
${\cal O}(g_{t,b}^2\,g^2 \gs^2)$ and ${\cal O}(g_{t,b}^2 \, \gpq
\gs^2)$ corrections. This term requires a computation of the
external-momentum dependence of the two-loop self-energy of the Higgs
boson, analogous to the one performed in refs.~\cite{Martin:2003qz,
  Martin:2003it, Martin:2004kr, Borowka:2014wla, Degrassi:2014pfa,
  Borowka:2018anu} for the ``diagrammatic'' case.
Finally, the last term on the r.h.s.~of eq.~(\ref{threeparts}) arises
from the fact that, while our calculation of the matching condition
for the quartic Higgs coupling $\lambda$ is performed in the $\drbar$
renormalization scheme assuming the field content of the MSSM, in the
EFT valid below the SUSY scale -- i.e., the SM -- $\lambda$ is interpreted
as an $\msbar$-renormalized quantity. Moreover, we find it convenient
to use $\msbar$-renormalized parameters of the SM also for the EW
gauge couplings entering the tree-level part of the matching
condition, see eq.~(\ref{looplam}), and for the top Yukawa coupling
entering the one-loop part (but not for the bottom Yukawa coupling, as
discussed in ref.~\cite{Bagnaschi:2017xid}). The corrections arising
from the change of renormalization scheme for $\lambda$ can in turn be
extracted from the two-loop self-energy diagrams for the Higgs boson,
those arising from the change of scheme and model for the EW gauge
couplings require a computation of the external-momentum dependence of
the two-loop self-energy of the $Z$ boson, while those arising from
the change of scheme and model for the top Yukawa coupling are easier
to obtain, being just the product of one-loop terms. In the rest of
this section we will describe in more detail our computation of each of the
three terms on the r.h.s.~of eq.~(\ref{threeparts}).

\vfill
\newpage

\subsection{1PI contributions}
\label{subsec:1PI}

The 1PI, two-loop contribution to the matching condition for the
quartic Higgs coupling can be expressed as
\beq
\label{effpot}
\Delta \lambda^{2\ell,{\rm 1{\scriptscriptstyle PI}}}
~ = ~ \frac12\,\left.
{\frac{\partial^4 \Delta V^{2\ell,\,\tilde q}}
{\partial^2H^\dagger\partial^2 H}}\,\right|_{H=0}~,
\eeq
where $\Delta V^{2\ell,\,\tilde q}$ denotes the contribution to the
MSSM scalar potential from two-loop diagrams involving the strong
gauge interactions of the squarks, and the derivatives are computed at
$H=0$ because we perform the matching between the MSSM and the SM in
the unbroken phase of the EW symmetry. Since the strong interactions
do not mix different types of squarks, we will now describe the
derivation of the stop contribution to $\Delta \lambda^{2\ell,{\rm
    1{\scriptscriptstyle PI}}}$, and later describe how to translate
it into the sbottom contribution and into the contributions of the
squarks of the first two generations.

It is convenient to start from the well-known expression for the stop
contribution of ${\cal O}(\gs^2)$ to the MSSM scalar potential in the
broken phase of the EW symmetry~\cite{Zhang:1998bm}, \bea
\label{v2as}
\Delta V^{2\ell,\,\tilde t} & = & \kappa^2\,\gs^2\,C_F N_c\,
\biggr\{ 2\,\tu\,I(\tu,\tu,0) + 2\,L(\tu,\g,\t)
- 4\,m_t\,\mg\,\sdt\,I(\tu,\g,\t) \nonumber\\
&& +\,\left(1-\frac{\sdt^2}2\right)\,J(\tu,\tu) 
+ \frac{\sdt^2}{2} J(\tu,\td)\;
+ \; \left[ \tul \leftrightarrow \tdl\,,\,
\sdt \rightarrow - \sdt\right] \biggr\}\,,
\eea
where $\kappa = 1/(16\pi^2)$ is a loop factor, $C_F=4/3$ and $N_c=3$
are color factors, the loop integrals $I(x,y,z)$, $L(x,y,z)$ and
$J(x,y)$ in eq.~(\ref{v2as}) are defined, e.g., in appendix D of
ref.~\cite{Degrassi:2009yq}, $\mg$ stands for the gluino mass, $\tul$
and $\tdl$ are the two stop-mass eigenstates, and $ \theta_{t}$
denotes the stop mixing angle. The latter is related to the top and
stop masses and to the left--right stop mixing parameter by
\beq
\label{s2t}
\sdt = \frac{2\, \mt\, X_t}{\tu-\td}~.
\eeq
We recall that $X_t = A_t - \mu\cot\beta$, where $A_t$ is the
trilinear soft SUSY-breaking Higgs--stop interaction term and $\mu$ is
the Higgs-higgsino mass parameter in the superpotential (in fact,
those two parameters enter our results only combined into $X_t$). To
compute the fourth derivative of the effective potential entering
eq.~(\ref{effpot}) we express the stop masses and mixing angle as
functions of a field-dependent top mass $m_t = \hat g_t\,|H|$, where
by $\hat g_t$ we denote~\footnote{We denote with a hat
  $\drbar$-renormalized couplings of the MSSM, and without a hat
  $\msbar$-renormalized couplings of the SM. However, in the two-loop
  part of the corrections the distinction between hatted and un-hatted
  couplings amounts to a higher-order effect, thus we will drop the
  hats there to reduce clutter.}  a SM-like Yukawa coupling related to
its MSSM counterpart $\hat y_t$ by $\hat g_t = \hat y_t\sin\beta$, and
of a field-dependent $Z$-boson mass $\MZ = \hgz \,|H|$, where we
define $\hgz^2 = (\hat g^2 + \hat g^{\prime\,2})/2$. We then obtain
\bea
\label{derivs}
\left.
\frac{\partial^4 \Delta V^{2\ell,\,\tilde t}}
{\partial^2H^\dagger\partial^2 H}\,\right|_{H=0}  &=&
\biggr[~  g_t^2\,\gz^2
\left( 2 \,V_{t\smallZ}^{(2)} \,+\, 12\,m_t^2 \,V_{tt\smallZ}^{(3)}
\,+\, 4\,m_t^4 \,V_{ttt\smallZ}^{(4)}\,+\, 3\,m_t^2 \,\MZ^2\, V_{tt\smallZ\smallZ}^{(4)}
\right)\nn\\
&& ~~+g_t^4\,\left( 2 \,V_{tt}^{(2)} \,+\, 4\,m_t^2 \,V_{ttt}^{(3)}
\,+\, m_t^4 \,V_{tttt}^{(4)}\right)\biggr]_{m_t,\MZ\,\rightarrow \,0}
~~+~~ \biggr[\,t \,\longleftrightarrow\, Z\biggr]~,
\eea
where the last term is obtained from the previous ones by swapping top
and $Z$, and we used the shortcuts
\beq
\label{shortcut}
V^{(k)}_{p_1\dots\, p_k} ~=~ 
\frac {d^k \Delta V^{2\ell,\,\tilde t}}{d m^2_{p_1}\dots\,d m^2_{p_k}}~,
~~~~~(p_i = t,Z)~.
\eeq
The terms proportional to $g_t^4$ in the second line of
eq.~(\ref{derivs}) give rise to the ${\cal O}(g_t^4\,\gs^2)$
contributions to $\Delta \lambda^{2\ell,{\rm 1{\scriptscriptstyle
      PI}}}$ already computed in ref.~\cite{Bagnaschi:2014rsa}, and we
will not consider them further. We will focus instead on the mixed
\qcdew\ contributions to $\Delta \lambda^{2\ell,{\rm
    1{\scriptscriptstyle PI}}}$ arising from the terms in
eq.~(\ref{derivs}) that are proportional to $g_t^2\,\gz^2$ and to
$\gz^4$. To obtain the derivatives with respect to $\mt$ and $\MZ$ of
the stop masses and mixing entering $\Delta V^{2\ell,\,\tilde t}$ we
exploit the relations
\beq
\label{derivst}
\frac{d m^2_{\tilde t_{1,2}}}{d\mt^2}~=~ 
1 \pm \frac{\sdt\,X_t}{2\,\mt}~,~~~~~~~~
\frac{d \sdt}{d\mt^2}~=~ 
\frac{\sdt\,\cdt^2}{2\,\mt^2}~,
\eeq
\beq
\label{derivsZ}
\frac{d m^2_{\tilde t_{1,2}}}{d\MZ^2}~=~ 
\frac{c_{2\beta}}{2}\,\left[d_\smallL^t + d_\smallR^t
  \pm \cdt\,(d_\smallL^t - d_\smallR^t)\right]~,~~~~~~~~
\frac{d \sdt}{d\MZ^2}~=~ 
-\frac{c_{2\beta}}{2}\,(d_\smallL^t - d_\smallR^t)\,\frac{\sdt^2\,\cdt}{\mt\,X_t}~,
\eeq
where
\beq
\label{dtLdtR}
d_\smallL^t = \frac12 - \frac23\, \sin^2\theta_\smallW~,~~~~~~~~
d_\smallR^t = \frac23\, \sin^2\theta_\smallW~,
\eeq
$\theta_\smallW$ being the Weinberg angle. After taking the required
derivatives of $\Delta V^{2\ell,\,\tilde t}$ with respect to~$\mt^2$
and $\MZ^2$, we use eq.~(\ref{s2t}) to make the dependence of
$\theta_t$ on $\mt$ explicit; we expand the function $\Phi(m^2_{\tilde
  t_{i}},\g,\mt^2)$ entering the loop integrals (see appendix D of
ref.~\cite{Degrassi:2009yq}) in powers of $\mt^2$; we take the limit
$|H|\rightarrow$~$0$, leading to $m_t,\MZ\,\rightarrow0$; and we
identify $\tul$ and $\tdl$ with the soft SUSY-breaking stop mass
parameters $m_{Q_3}$ and $m_{U_3}$. We remark that, differently from
the ${\cal O}(g_t^4\,\gs^2)$ contributions computed in
ref.~\cite{Bagnaschi:2014rsa} and the ${\cal O}(g_t^6)$ contributions
computed in ref.~\cite{Bagnaschi:2017xid}, the mixed \qcdew\ 
contributions to $\Delta \lambda^{2\ell,{\rm 1{\scriptscriptstyle
      PI}}}$ computed here do not contain infrared divergences that
need to be canceled out in the matching of the quartic Higgs coupling
between MSSM and SM. Also, the terms proportional to $ g_t^2\gz^2$ and
to $\gz^4$ in eq.~(\ref{derivs}) that involve more than two
derivatives of the two-loop effective potential vanish directly when
we take the limit $|H|\rightarrow 0$, thus the mixed \qcdew\ 
contributions to $\Delta \lambda^{2\ell,{\rm 1{\scriptscriptstyle
      PI}}}$ can be related as usual to the corresponding two-loop
corrections to the Higgs mass.


In order to obtain the contributions to $\Delta \lambda^{2\ell,{\rm
    1{\scriptscriptstyle PI}}}$ from diagrams involving sbottoms, it
is sufficient to perform the replacements $g_t\rightarrow g_b\,$,~
$X_t\rightarrow X_b\,$,~ $m_{U_3}\rightarrow m_{D_3}$~ and
$d_{\smallL,\smallR}^t\rightarrow d_{\smallL,\smallR}^b$~ in the
contributions from diagrams involving stops, with
\beq
\label{dbLdbR}
d_\smallL^b = -\frac12 + \frac13\, \sin^2\theta_\smallW~,~~~~~~~~
d_\smallR^b = -\frac13\, \sin^2\theta_\smallW~.
\eeq
Finally, we recall that our calculation neglects the Yukawa couplings
of the first two generations. The contributions to $\Delta
\lambda^{2\ell,{\rm 1{\scriptscriptstyle PI}}}$ from diagrams
involving up-type (or down-type) squarks of the first two generations
can be obtained by setting $g_t =0$ (or $g_b =0$) in the contributions
from diagrams involving stops (or sbottoms), and replacing the soft
SUSY-breaking stop (or sbottom) mass parameters with those of the
appropriate generation.

\vfill
\newpage

The result for $\Delta \lambda^{2\ell,{\rm 1{\scriptscriptstyle PI}}}$
with full dependence on all of the input parameters is lengthy and not
particularly illuminating, and we make it available upon request --
together with all of the other corrections computed in this paper --
in electronic form. We show here a simplified result valid in the
limit in which all squark masses ($m_{Q_i},\, m_{U_i},\, m_{D_i},\,$ with
$i=1,2,3$) as well as the gluino mass $\mg$ are set equal to a common
SUSY scale $\MS$. In units of $\kappa^2\,\gs^2\,C_FN_c$, we find
\bea
\Delta \lambda^{2\ell,{\rm 1{\scriptscriptstyle PI}}} &=&
\frac34 \left(g^4 + \frac{11}{9}\,g^{\prime\,4}\right)\,\cdbe^2\nn\\[1mm]
&-&\left(g^2 + g^{\prime\,2}\right)\,\cdbe\,
\left[\,g_t^2\left(1+ \ln^2\frac{\MS^2}{Q^2} 
  - 2\,\frac{X_t}{\MS}\,\ln\frac{\MS^2}{Q^2} -\frac{X_t^2}{\MS^2} \right)
  ~-~~(t\rightarrow b)~\right]~,
\eea
where $(t\rightarrow b)$ denotes terms obtained from the previous ones
within square brackets via the replacements $g_t\rightarrow g_b$ and
$X_t \rightarrow X_b$\,. We remark in passing that there are no
contributions of ${\cal O}(g^2\,g^{\prime\,2}\,\gs^2)$, because the
corresponding diagrams involve the trace of the generators of the
$SU(2)$ gauge group.

\subsection{WFR contributions}
\label{subsec:WFR}

Differently from the case of the ``gaugeless'' corrections computed in
refs.~\cite{Bagnaschi:2014rsa, Bagnaschi:2017xid}, where the quartic
Higgs coupling can be considered vanishing at tree level, the mixed
\qcdew\  corrections include a contribution in which the tree-level
coupling is combined with the two-loop WFR of the Higgs field. This
reads
\beq
\label{dlamWFR}
\Delta \lambda^{2\ell,{\rm {\scriptscriptstyle WFR}}} ~=~
-\frac12\,(g^2 + g^{\prime\,2})\,\cdbe^2\,
\left( \left.\frac{d\,\hat \Pi_{hh}^{2\ell,\,\tilde q}}{dp^2}\,\right|_{H=0}
\!+~ \dWFR~\right), 
\eeq
where $\hat \Pi_{hh}^{2\ell,\,\tilde q}$ denotes the contribution to
the renormalized Higgs-boson self-energy\,\footnote{For the
  self-energies of both the Higgs boson and the $Z$ boson, we adopt in
  this paper the sign convention according to which $m^2_{\rm pole}=
  m^2_{\rm run} - \Pi(m^2)$\,. The WFR for the Higgs boson is then $Z_h
  = 1 - d\Pi_{hh}(p^2)/dp^2$\,. Note that this is the opposite of the
  sign convention adopted in refs.~\cite{Martin:2003qz, Martin:2003it,
    Martin:2004kr} and \cite{Braathen:2018htl}.} from two-loop
diagrams involving the strong gauge interactions of the squarks, and
the derivative is taken with respect to the external momentum
$p^2$. In this case the notation $H=0$ means that, after having taken
the derivative, we take the limit $v\rightarrow 0$. This implies
$m_q,\MZ\rightarrow 0$ as well as $p^2\rightarrow 0$\, (because $p^2$
is ultimately set to $2\lambda v^2$). Finally, the shift $\dWFR$ stems
from the matching of the one-loop WFR, and will be discussed below.

The relevant two-loop self-energy diagrams are generated with
\texttt{FeynArts}\cite{Hahn:2000kx}, using a modified version of the
original MSSM model file~\cite{Hahn:2001rv} that implements the QCD
interactions in the background field gauge.  The color factors are
simplified with a private package and the Dirac algebra is handled by
\texttt{TRACER}~\cite{Jamin:1991dp}.  In order to obtain a result
valid in the limit $v\rightarrow 0$, we performed an asymptotic
expansion of the self-energy in the heavy superparticle masses
analogous to the one described in section 3 of
ref.~\cite{Degrassi:2010eu}. As a useful cross-check of our result, we
verified that it agrees with the one that can be obtained by taking
appropriate limits in the explicit analytic formulae for the
Higgs-boson self-energy given in refs.~\cite{Martin:2003qz,
  Martin:2003it, Martin:2004kr}.

Our result for the derivative of the two-loop self-energy with full
dependence on all of the input parameters is in turn made available
upon request. In the simplified scenario of degenerate superparticle
masses we find, in units of $\kappa^2\,\gs^2\,C_FN_c$,
\bea
\label{dPidp2}
\left.\frac{d\,\hat \Pi_{hh}^{2\ell,\,\tilde q}}{dp^2}\,\right|_{H=0}
&=& g_t^2\,\left[\,
  \frac{13}{4} ~+~ 3\, \ln\frac{\MS^2}{Q^2} ~+\, \ln^2\frac{\MS^2}{Q^2}
  ~-~ \frac{2\,X_t}{3\,\MS}\left( 4+\ln\frac{\MS^2}{Q^2}\right)
  \,+~ \frac{7\,X_t^2}{6\,\MS^2}
\right.\nn\\[2mm]
&&~~~~~~~\left.+~2\,\left(\ln\frac{\MS^2}{Q^2} \,-\, \frac{X_t}{\MS}\right)
  \left(1\,-\,\ln\frac{-p^2}{Q^2}\right)~\right]~~+~~ (t\rightarrow b)~.
\eea  
We remark that there are no contributions proportional to the EW gauge
couplings. This is due to the fact that, in the limit of unbroken EW
symmetry, the $D$-term-induced ${\cal O}(g^2)$ and ${\cal
  O}(g^{\prime\,2})$ couplings of the Higgs boson to squarks enter
only self-energy diagrams that do not depend on the external
momentum. We also remark that the derivative of the two-loop
self-energy has logarithmic infra-red (IR) divergences in the limit
$p^2\rightarrow 0$, see the last term within square brackets in the
second line of eq.~(\ref{dPidp2}). These divergences cancel out in the
matching of the Higgs-boson WFR between the MSSM and the SM. Indeed,
in the limit $v\rightarrow0$ the one-loop contribution of top and
bottom quarks to the derivative of the self-energy, which is present
both above and below the matching scale, reads
\beq
\label{dPidp2oneloop}
\left.\frac{d\,\hat \Pi_{hh}^{1\ell,\,q}}{dp^2}\,\right|_{H=0}
=~~ \kappa\,N_c \,(g_t^2 + g_b^2)\,
\left(1\,-\,\ln\frac{-p^2}{Q^2}\right)~.
\eeq
However, the top and bottom Yukawa couplings must be interpreted as
the ones of the MSSM above the matching scale, and the ones of the SM
below the matching scale. Thus, in the matching between the MSSM and
the SM the derivative of the two-loop Higgs self-energy receives the
shift
\beq
\label{shiftWFR}
\dWFR ~=~ 2\,\kappa\,N_c \,\left(g_t^2 \,\Delta g_t^{\tilde t,\,\gs^2}
+ g_b^2 \,\Delta g_b^{\tilde b,\,\gs^2}\right)\,
\left(1\,-\,\ln\frac{-p^2}{Q^2}\right)~,
\eeq
where $\Delta g_t^{\tilde t,\,\gs^2}$ denotes the one-loop, ${\cal
  O}(\gs^2)$ contribution from diagrams involving stops to the
difference between the MSSM coupling $\hat g_t$ and the SM coupling
$g_t$,
\beq
\label{deltagtsusy}
\Delta g_t^{\tilde t,\,\gs^2} ~=~ - \kappa\,\gs^2\,C_F\,\left[
\ln\frac{\mg^2}{Q^2} 
\,+\, \wt F_6\left(\frac{m_{Q_3}}{\mg}\right)
\,+\, \wt F_6\left(\frac{m_{U_3}}{\mg}\right)
\,-\, \frac{X_t}{\mg}\,
\wt F_9\left(\frac{m_{Q_3}}{\mg},\frac{m_{U_3}}{\mg}\right)\right]~,
\eeq
and the analogous shift in the bottom Yuakwa coupling, $\Delta
g_b^{\tilde b,\,\gs^2}$, can be obtained from eq.~(\ref{deltagtsusy})
with the replacements $m_{U_3}\rightarrow m_{D_3}$ and $X_t\rightarrow
X_b$. The loop functions $\wt F_6(x)$ and $\wt F_9(x,y)$ are defined
in the appendix A of ref.~\cite{Bagnaschi:2014rsa}. Using the limits
$\wt F_6(1)=0$ and $\wt F_9(1,1)=1$ it is easy to see that, for
degenerate superparticle masses, the shifts in eq.~(\ref{shiftWFR})
cancel out entirely the terms within square brackets in the second
line of eq.~(\ref{dPidp2}). We have of course checked that the
cancellation of the IR divergences in the matching holds even when we
retain the full dependence on all of the relevant superparticle
masses.

\vfill

\subsection{Contributions arising from the definitions of the couplings}
\label{subsec:RSC}

The third contribution to the mixed \qcdew\  correction to the quartic
Higgs coupling in eq.~(\ref{threeparts}), which we denoted as $\Delta
\lambda^{2\ell,\,{\rm {\scriptscriptstyle RS}}}$, collects in fact
three separate contributions arising from differences in the
renormalization scheme used for the couplings of the MSSM and for
those of the EFT valid below the matching scale (i.e., the SM):
\beq
\label{dlambdascheme}
\Delta\lambda^{2\ell,\,{\rm {\scriptscriptstyle RS}}} ~=~
\Delta\lambda_\lambda^{2\ell,\,{\rm {\scriptscriptstyle RS}}} ~+~
\Delta\lambda_{g}^{2\ell,\,{\rm {\scriptscriptstyle RS}}} ~+~
\Delta\lambda_{g_t}^{2\ell,\,{\rm {\scriptscriptstyle RS}}} ~.
\eeq

The first contribution in the equation above stems from the fact that
supersymmetry provides a prediction for the $\drbar$-renormalized
quartic Higgs coupling, whereas we interpret the parameter $\lambda$
in the EFT as renormalized in the $\msbar$ scheme. The difference
between $\lambda^{\smalldrbar}$ and $\lambda^{\smallmsbar}$ contains
terms of ${\cal O}(\lambda\,g_t^2\,\gs^2)$ and ${\cal
  O}(\lambda\,g_b^2\,\gs^2)$, arising from the dependence on the
regularization method of the two-loop quark--gluon contributions to the
Higgs WFR, which translate to mixed \qcdew\  terms when $\lambda$ is
replaced by its tree-level MSSM prediction. In the limit $v\rightarrow
0$ we find
\beq
\label{WFRreg}
\left.\frac{d\,\hat \Pi_{hh}^{2\ell,\,q}}{dp^2}\,\right|_{H=0}^{\smalldred} \!-~
\left.\frac{d\,\hat \Pi_{hh}^{2\ell,\,q}}{dp^2}\,\right|_{H=0}^{\smalldreg}
\!=~2\,\kappa^2\,\gs^2\,(g_t^2+g_b^2)\,C_F\,N_c\,
\left(1\,-\,\ln\frac{-p^2}{Q^2}\right)
-~ \frac12 \,\kappa^2\,\gs^2\,(g_t^2+g_b^2)\,C_F\,N_c~,
\eeq
where DRED and DREG stand for dimensional reduction and dimensional
regularization, respectively. This is again in agreement with the
result that can be obtained by taking the appropriate limits in the
analytic formulae of refs.~\cite{Martin:2003qz, Martin:2003it,
  Martin:2004kr}. The first term on the r.h.s.~of eq.~(\ref{WFRreg})
contains an IR divergence for $p^2\rightarrow 0$, but that term
cancels out in the matching between MSSM and SM when the top and
bottom Yukawa couplings entering the one-loop quark contribution to
the WFR, see eq.~(\ref{dPidp2oneloop}), are translated from the
$\drbar$ scheme to the $\msbar$ scheme according to
\beq
\label{dgtreg}
g_q^{\smalldrbar} ~=~ g_q^{\smallmsbar}
\,\left(1+\Delta g_q^{{\scriptscriptstyle {\rm reg}},\,\gs^2}\right)\,,
~~~~~~{\rm with}~~~~~~\Delta g_q^{{\scriptscriptstyle {\rm reg}},\,\gs^2}
       ~=\, -\kappa\,\gs^2\,C_F~.
\eeq
The surviving term on the r.h.s.~of eq.~(\ref{WFRreg}) then leads to
the following correction to the quartic Higgs coupling:
\beq
\Delta\lambda_\lambda^{2\ell,\,{\rm {\scriptscriptstyle RS}}} ~=~
\frac14 \,\kappa^2\,\gs^2\,C_F\,N_c\,(g_t^2+g_b^2)\,(g^2+g^{\prime\,2})\,\cdbe^2~.
\eeq

\bigskip

Supersymmetry connects the tree-level quartic Higgs coupling to the
$\drbar$-renormalized EW gauge couplings of the MSSM. Since we choose
instead to express the tree-level part of the matching condition for
$\lambda$, see eq.~(\ref{looplam}), in terms of $\msbar$-renormalized
couplings of the SM, the threshold correction $\Delta\lambda^{2\ell}$
receives an additional shift, i.e.~the second contribution in
eq.~(\ref{dlambdascheme}). The relation between the two sets of EW
gauge couplings reads
\beq
\hat g^2 + \hat g ^{\prime\,2} ~=~
(g^2 + g ^{\prime\,2})~\left( 1 ~+~ \ldots ~+~
\left.\frac{d\,\hat\Pi_{\smallZ\smallZ}^{2\ell,\,q}}{dp^2}\,\right|_{H=0}^{\smalldred}\!-~
\left.\frac{d\,\hat\Pi_{\smallZ\smallZ}^{2\ell,\,q}}{dp^2}\,\right|_{H=0}^{\smalldreg}\!+~
\left.\frac{d\,\hat\Pi_{\smallZ\smallZ}^{2\ell,\,\tilde q}}{dp^2}\,\right|_{H=0}~\right)~,
\eeq
where the ellipsis denotes one- and two-loop terms that are not of
${\cal O}(g^4\,\gs^2)$ or ${\cal O}(g^{\prime\,4}\,\gs^2)$.  We denote
by $\hat\Pi_{\smallZ\smallZ}^{2\ell,\,q}$ the two-loop quark--gluon
contribution to the transverse part of the renormalized $Z$-boson
self-energy, computed either in dimensional reduction or in
dimensional regularization, and by
$\hat\Pi_{\smallZ\smallZ}^{2\ell,\,\tilde q}$ the contribution from
two-loop diagrams involving the strong gauge interactions of the
squarks.\footnote{The fact that these mixed \qcdew\  corrections should
  depend only on the quark and squark contributions to the gauge-boson
  self-energy can be easily inferred by considering the
  renormalization of the gauge couplings of the leptons.} The notation
$H=0$ means again the limit $v\rightarrow 0$, which in this case can
be obtained by Taylor expansion in $m_q^2$ and $p^2$, since both the
squark contribution and the DRED--DREG difference of the quark
contribution are free of IR divergences. In units of
$\kappa^2\,\gs^2\,C_FN_c$\,, we find
\bea
\label{shiftEW}
\Delta\lambda_g^{2\ell,\,{\rm {\scriptscriptstyle RS}}} &=&
\frac{\cdbe^2}{4}\,\left\{-\frac{3\,g^4}{2}
~-~\frac{11\,g^{\prime\,4}}{6}
\phantom{\sum_{i=1}^3}\right.\nn\\
  &&~~~~~~~- \left.
  \sum_{i=1}^3 \left[\,g^4\,F(m^2_{Q_i},\mg^2) \,+\, \frac{~g^{\prime\,4}}{9}
  \biggr(F(m^2_{Q_i},\mg^2) + 8\,F(m^2_{U_i},\mg^2)+2\,F(m^2_{D_i},\mg^2)
  \biggr)\right]\right\},\nn\\
\eea
where the two terms within curly brackets in the first line account
for the $\drbar$--$\msbar$ conversion of $g^2$ and $g^{\prime\,2}$,
respectively, whereas the two terms in the second line (where the sum
runs over three squark generations) account for their MSSM--SM threshold
correction. The function $F(\msqq,\mg^2)$ is defined as
\beq
\label{funcF}
F(\msqq,\mg^2) ~=~
\frac{1}{12}\left(7+\frac{8\,\mg^2}{\msqq}\right)
\,+~\frac{2\,\mg^4}{3\,\msqq\,(\msqq-\mg^2)}\,\ln\frac{\mg^2}{\msqq}
~+\,\left(1-\frac{2\,\mg^2}{3\,\msqq}\right)\,\ln\frac{\msqq}{Q^2}~.
\eeq
We checked that the explicit renormalization-scale dependence of the
${\cal O}(g^4\,\gs^2)$ and ${\cal O}(g^{\prime\,4}\,\gs^2)$ threshold
corrections to $g^2$ and $g^{\prime\,2}$ is consistent with what can
be inferred from the difference between their $\beta$-functions in the
SM~\cite{Machacek:1983tz} and in the MSSM~\cite{Jones:1974pg,
  Jones:1983vk, Martin:1993zk}.

Finally, the third contribution in eq.~(\ref{dlambdascheme}) arises
from the fact that we choose to express the ${\cal O}(g^2\,g_t^2)$ and
${\cal O}(g^{\prime\,2}\,g_t^2)$ terms in the one-loop threshold
correction to the quartic Higgs coupling in terms of the
$\msbar$-renormalized top Yukawa coupling of the SM. The resulting
shift in the two-loop correction reads
\beq
\label{shiftgt}
\Delta\lambda_{g_t}^{2\ell,\,{\rm {\scriptscriptstyle RS}}} ~=~
2\,\left(\,\Delta g_t^{\tilde t,\,\gs^2}
+\,\Delta g_q^{{\scriptscriptstyle {\rm reg}},\,\gs^2}\right)
\,\Delta\lambda^{1\ell,\, \tilde t_{\rm \,\smallEW}}~,
\eeq
where $\Delta g_t^{\tilde t,\,\gs^2}$ and $\Delta
g_q^{{\scriptscriptstyle {\rm reg}},\,\gs^2}$ are the one-loop, ${\cal
  O}(g_s^2)$ shifts given in eqs.~(\ref{deltagtsusy}) and
(\ref{dgtreg}), respectively, and
\bea
\Delta\lambda^{1\ell,\, \tilde t_{\rm \,\smallEW}} \!&=&
\kappa\,N_c\,g_t^2\,\cdbe
\left\{ \frac12 \left(g^2-\frac{g^{\prime\,2}}{3}\right)\ln\frac{m_{Q_3}^2}{Q^2}
  \,+\,\frac{2\,g^{\prime\,2}}{3}\,\ln\frac{m_{U_3}^2}{Q^2}\right.\nn\\[2mm]
  &&~~~~~~~~~~~~~~~~\left.
  +\frac{X_t^2}{4\,m_{Q_3}m_{U_3}}\,\left[g^{\prime\,2}\,\wt F_3(\xqu)
    \,+\, g^2\,\wt F_4(\xqu)
    \,-\, \frac{\cdbe}{3}\,(g^2 + g^{\prime\,2})\,\wt F_5(\xqu)
    \right]\right\}~,\nn\\
\eea
where $\xqu=m_{Q_3}/m_{U_3}$ and the loop functions $\wt F_3(x)$, $\wt
F_4(x)$ and $\wt F_5(x)$ are defined in the appendix A of
ref.~\cite{Bagnaschi:2014rsa}. Note that all three functions are equal
to 1 for $x=1$.


\vfill

\subsection{Combining all contributions}

We now provide a result that combines all of the contributions
discussed in the previous sections, valid in the limit of degenerate
superparticle masses $m_{Q_i}=m_{U_i}=m_{D_i}=\mg=\MS$. In units of
$\kappa^2\,\gs^2\,C_FN_c$\,, we find
\bea
\label{dlambdatotal}
\Delta \lambda^{2\ell,\,{\rm {\scriptscriptstyle QCD\mbox{-}EW}}} &=&
\cdbe^2\vast\{-\frac{1}{16}\,\left(g^4+\frac{11}{9}\,g^{\prime\,4}\right)
\left(1+4\,\ln\frac{\MS^2}{Q^2}\right)\nn\\[2mm]
&&~~~~~~+(g^2 + g^{\prime\,2})\left\{\,
  (g_t^2+ g_b^2)\left(\,-\frac{11}{8}-\frac32\,\ln\frac{\MS^2}{Q^2}
  -\frac12\,\ln^2\frac{\MS^2}{Q^2}\right)
  \right.\nn\\[2mm]
  &&~~~~~~~~~~~~~~~~~~~~~~~+g_t^2\left[\,
  \frac{X_t}{3\,\MS}\left(4+\ln\frac{\MS^2}{Q^2}\right)    
  -\frac{X_t^2}{12\,\MS^2}\left(5-2\,\ln\frac{\MS^2}{Q^2}\right)
  -\frac{X_t^3}{6\,\MS^3}~\right]\nn\\[2mm]
  &&~~~~~~~~~~~~~~~~~~~~~~~\left. +\,g_b^2\left[\, 
    \frac{X_b}{3\,\MS}\left(4+\ln\frac{\MS^2}{Q^2}\right)
    -\frac{7\,X_b^2}{12\,\MS^2}~\right]\right\}\vast\}\nn\\[2mm]
  &+&\cdbe\,(g^2 + g^{\prime\,2})\left\{\,
  g_t^2\left[\,-1-\ln\frac{\MS^2}{Q^2}-2\,\ln^2\frac{\MS^2}{Q^2}
    \right.\right.\nn\\[2mm]
    &&~~~~~~~~~~~~~~~~~~~~~~~~~
    \left. +3\,\frac{X_t}{\MS}\,\ln\frac{\MS^2}{Q^2}
    +\frac{X_t^2}{2\,\MS^2}\left(1-\ln\frac{\MS^2}{Q^2}\right)
    +\frac{X_t^3}{2\,\MS^3}\,\right]\nn\\[2mm]
  &&~~~~~~~~~~~~~~~~~~\left.+g_b^2\left[\,1 + \ln^2\frac{\MS^2}{Q^2}
    -2\,\frac{X_b}{\MS}\,\ln\frac{\MS^2}{Q^2}
    -\frac{X_b^2}{\MS^2}\right]~\right\}~~.
\eea

\bigskip

As a non-trivial check of our final result, we verified that by taking
the derivative of the r.h.s.~of eq.~(\ref{looplam}) with respect to
$\ln Q^2$ we can recover the ${\cal O}(\lambda\,g_t^2\,\gs^2)$ and
${\cal O}(\lambda\,g_b^2\,\gs^2)$ terms of the $\beta$-function for
the quartic Higgs coupling of the SM~\cite{Machacek:1984zw}:
\beq
\label{rgelam}
\frac{d\lambda}{d\ln Q^2} ~\supset~ 40\,\kappa^2\,\lambda
\,\gs^2\,(g_t^2+g_b^2)~.
\eeq
To this effect, we must combine the explicit scale dependence of our
result for $\Delta \lambda^{2\ell,\,{\rm {\scriptscriptstyle
      QCD\mbox{-}EW}}}$ with the implicit scale dependence of the
parameters that enter the tree-level and one-loop parts of the
matching condition (for the squark contributions to the latter, see
the appendix of ref.~\cite{Bagnaschi:2017xid}). In particular, we need
the terms that involve the strong gauge coupling in the two-loop
$\beta$-functions of the EW gauge couplings~\cite{Machacek:1983tz},
\beq
\label{rgeEW}
\frac{dg^2}{d\ln Q^2} ~\supset~ 12\,\kappa^2\,g^4\,\gs^2~,~~~~~~~~
\frac{dg^{\prime\,2}}{d\ln Q^2} ~\supset~
\frac{44}{3}\,\kappa^2\,g^{\prime\,4}\,\gs^2~,
\eeq
in the two-loop $\beta$-function of
$\cdbe$~\cite{Sperling:2013xqa},
\beq
\frac{d\cdbe}{d\ln Q^2} ~\supset~ (3\,\kappa + 16\,\kappa^2\,\hat \gs^2)\,
\left[\hat g_t^2-\hat g_b^2 \,+\, \cdbe\,(\hat g_t^2 + \hat g_b^2)\,\right]~,
\eeq
and in the one-loop $\beta$-functions of the parameters that enter
$\Delta \lambda^{1\ell}$,
\beq
\frac{d \,(m^2_{Q_i},m^2_{U_i},m^2_{D_i})}{d\ln Q^2} \,~\supset~
-\frac{16}{3}\,\kappa\,\hat \gs^2\,\mg^2~,~~~~~~~~
\frac{d \,(X_t,X_b)}{d\ln Q^2} ~\supset~
\frac{16}{3}\,\kappa\,\hat \gs^2\,\mg~,
\eeq

\beq
\label{rgeYuk}
\frac{d g_t^2}{d\ln Q^2} ~\supset~
-8\,\kappa\,g_t^2\,\gs^2~,~~~~~~~~
\frac{d \hat g_b^2}{d\ln Q^2} ~\supset~
-\frac{16}{3}\,\kappa\,\hat g_b^2\,\hat \gs^2~.
\eeq
Note that in eqs.~(\ref{rgeEW})--(\ref{rgeYuk}) we distinguish the
(hatted) $\drbar$-renormalized couplings of the MSSM from the
(unhatted) $\msbar$-renormalized couplings of the SM (however, the
distinction is irrelevant for the strong gauge coupling, which enters
only the two-loop part of the calculation). Finally, to recover
eq.~(\ref{rgelam}) we need to convert the remaining MSSM Yukawa
couplings in the one-loop part of the derivative of $\lambda$ into
their SM counterparts, and exploit in the two-loop part the tree-level
MSSM relation to replace $(g^2+g^{\prime\,2})\,\cdbe^2$ with
$4\lambda$\,.

\section{The EW-scale determination of the Higgs mass}
\label{sec:weakscale}

A consistent determination of the Higgs mass in the EFT approach
requires that the relation between the pole Higgs mass and the
$\msbar$-renormalized quartic Higgs coupling at the EW scale be
computed at the same perturbative order in the various SM couplings as
the SUSY-scale threshold correction $\Delta\lambda$. In particular,
the ``gaugeless'' calculation of refs.~\cite{Bagnaschi:2014rsa,
  Bagnaschi:2017xid} requires a full determination of the one-loop
corrections to the Higgs mass at the EW scale, combined with the
two-loop corrections obtained in the limit
$g=g^{\prime}=\lambda=0$. Denoting the scale at which we perform the
calculation of the Higgs mass as $\qew$, this approximation implies
\bea
\label{2loopSMgl}
m_h^2 &=&
\frac{\lambda(\qew)}{\sqrt{2}\,G_F}\,\left[1-\delta^{1\ell}(\qew)\right]\,\nn\\
&+& 8\,\kappa^2\,C_F\,N_c\,g_t^2\,\gs^2\,m_t^2\left(3\,\ell_t^2 + \ell_t\right)
\,-~ 2\,\kappa^2\,N_c\,g_t^4\,m_t^2\left(9\,\ell_t^2
- 3\,\ell_t + 2 + \frac{\pi^2}{3}
\right)~,
\eea
where $G_F$ is the Fermi constant, $\delta^{1\ell}$ is the one-loop
correction first computed in ref.~\cite{Sirlin:1985ux}, and $\ell_t
=$~$\ln(\mt^2/\qew)$. The two-loop terms in the second line of
eq.~(\ref{2loopSMgl}) are taken from ref.~\cite{Degrassi:2012ry}. Note
that the form of the ${\cal O}(g_t^4\,\mt^2)$ terms implies that the
top-quark contribution to $\delta^{1\ell}$ is expressed in terms of
$\msbar$-renormalized top and Higgs masses. Also, note that
eq.~(\ref{2loopSMgl}) omits for conciseness all corrections involving
the bottom and tau Yukawa couplings, which in the SM are greatly
suppressed with respect to their top-only counterparts.

In the SUSY-scale calculation of $\Delta\lambda^{2\ell}$ described in
section \ref{sec:matching} we go beyond the gaugeless limit, and we
include contributions that involve both the EW gauge couplings and the
strong gauge coupling. Strictly speaking, to match the accuracy of
that calculation at the EW scale we only need to replace the ${\cal
  O}(g_t^2\,\gs^2\,\mt^2)$ terms in eq.~(\ref{2loopSMgl}) with the
complete contributions arising from two-loop diagrams involving quarks
and gluons, retaining the dependence on the external momentum in the
Higgs self-energy. Explicit formulae for those contributions are
provided in ref.~\cite{Bezrukov:2012sa}. However, the full two-loop
contributions to the relation between the pole Higgs mass,
$\lambda(\qew)$ and $G_F$ in the SM are also
available~\cite{Buttazzo:2013uya, Kniehl:2015nwa,
  Martin:2019lqd}. These contributions can in principle be implemented
in our EFT calculation, to prepare the ground for the eventual
completion of the NNLL resummation of the large logarithmic
corrections. We stress that the inclusion at the EW scale of two-loop
corrections whose counterparts are still missing at the SUSY scale
cannot be claimed to improve the overall accuracy of the calculation,
but it does not degrade it either. Indeed, in the EFT approach the
EW-scale and SUSY-scale sides of the calculation are separately free
of log-enhanced terms, and the inclusion of additional pieces in only
one side does not entail the risk of spoiling crucial cancellations
between large corrections.

The results of the full two-loop calculation of
ref.~\cite{Buttazzo:2013uya} were made public in the form of
interpolating formulae. In particular, the relation between the Higgs
quartic coupling and the pole Higgs and top masses
reads~\footnote{Note that our normalization for $\lambda$ differs from
  the one in refs.~\cite{Buttazzo:2013uya,Kniehl:2015nwa,
    Martin:2019lqd} by a factor 2.}
\beq
\label{interp}
\lambda(\qew=\mt) ~=~ 0.25208
\,+\, 0.00412\,\left(\frac{\mh}{{\rm GeV}} - 125.15\right)
  \,-\, 0.00008\,\left(\frac{\mt}{{\rm GeV}} - 173.34\right)
    \,\pm\, 0.00060_{\rm \,th}~,
\eeq
where the remaining input parameters -- namely, the gauge-boson
masses, $G_F$ and $\gs(\MZ)$ -- are set to the central values listed
in ref.~\cite{Buttazzo:2013uya}. Eq.~(\ref{interp}) can be exploited
to treat the measured value of the Higgs mass as an input parameter:
it is then possible to evolve $\lambda$ to the SUSY scale using the
RGEs of the SM, and use the threshold condition in eq.~(\ref{looplam})
to determine one of the MSSM parameters (e.g., a common mass scale for
the stops, or the stop mixing parameter $X_t\,$, or $\tan\beta$) as a
function of the others. In alternative, eq.~(\ref{interp}) can be
inverted to predict the Higgs mass starting from a full set of MSSM
parameters, using the value of $\lambda(\mt)$ obtained by evolving the
$\lambda(\MS)$ computed in eq.~(\ref{looplam}) down to the EW
scale. In the latter approach, a phenomenological analysis of the MSSM
may well encounter points of the parameter space in which the
prediction for the Higgs mass is several GeV away from the measured
value. It is then legitimate to wonder about the range of validity of
the linear interpolation involved in eq.~(\ref{interp}), which was
obtained in a pure-SM context with the (small) uncertainty of the
Higgs mass measurement in mind.

\begin{figure}[t]
\begin{center}
\vspace*{-1cm}
\includegraphics[width=13cm]{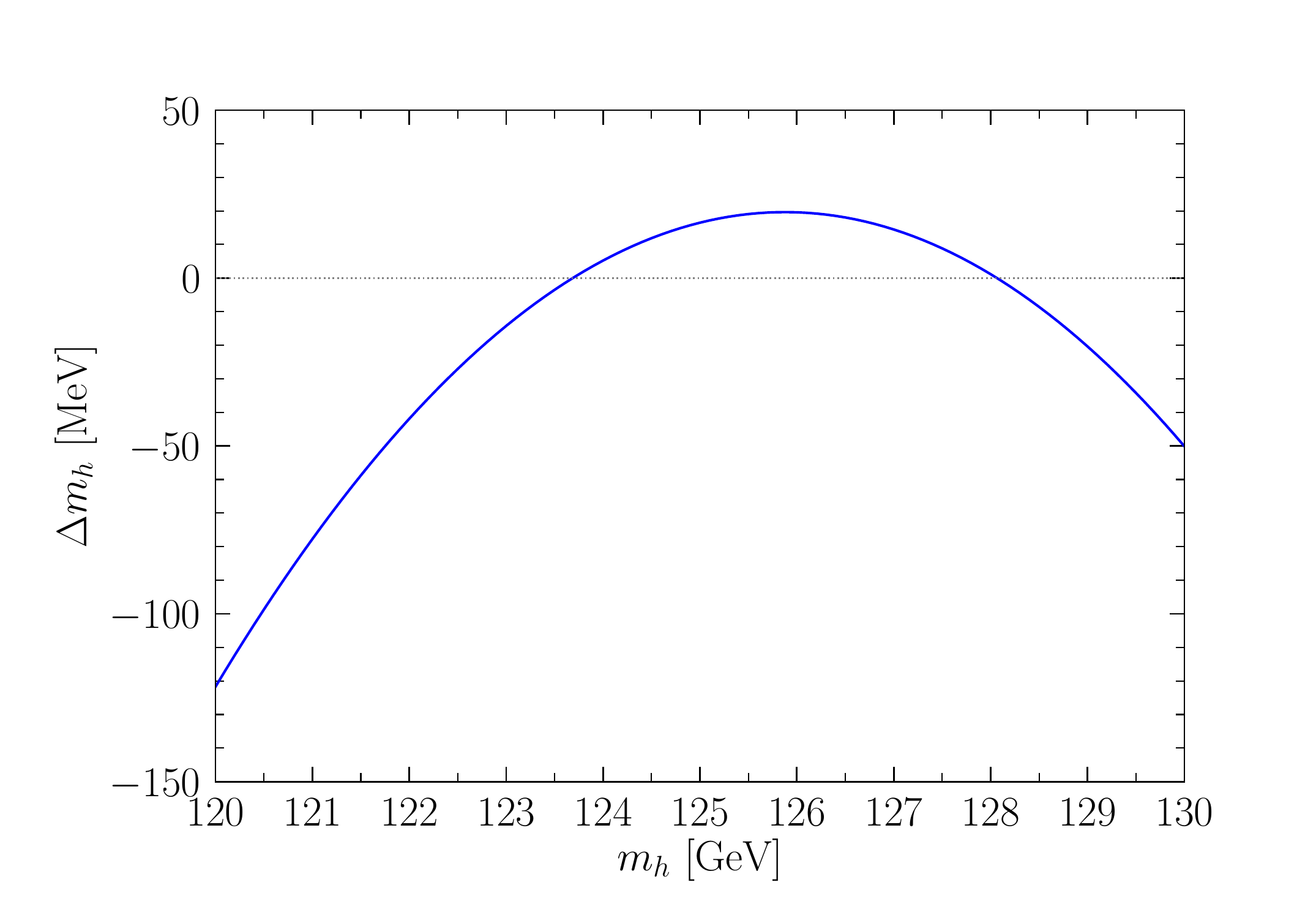}
  \caption{\em Difference (in MeV) between the Higgs mass given as
    input to the code {\tt mr} and the Higgs mass obtained by
    inserting in eq.~(\ref{interp}) the value of $\lambda(\mt)$
    computed by {\tt mr}, as a function of the input Higgs mass.}
  \label{fig:buttvsmr}
\vspace*{-1mm}
\end{center}
\end{figure}

To test eq.~(\ref{interp}), we compare its predictions against those
of the independent two-loop calculation presented in
ref.~\cite{Kniehl:2015nwa}. The latter is made available in the public
code {\tt mr}~\cite{Kniehl:2016enc}, which computes the
$\msbar$-renormalized parameters of the SM Lagrangian from a set of
physical observables that includes $G_F$ and the pole masses of the
Higgs and gauge bosons and of the top and bottom quarks. We start from
an input value for the pole Higgs mass ranging between $120$~GeV and
$130$~GeV, feed it into {\tt mr}, then insert the value of
$\lambda(\mt)$ computed by {\tt mr} into eq.~(\ref{interp}) to obtain
a new prediction for $\mh$ according to the calculation of
ref.~\cite{Buttazzo:2013uya}.  The remaining input parameters are
fixed to the central values considered in
ref.~\cite{Buttazzo:2013uya}. In figure~\ref{fig:buttvsmr} we plot the
difference between the initial and final values of $\mh$, which can be
taken as a measure of the discrepancy between the two calculations, as
a function of the initial value. In the vicinity of $\mh = 125$~GeV,
where the interpolation involved in eq.~(\ref{interp}) can be expected
to be accurate, the two values of $\mh$ differ by about $20$~MeV,
i.e.~by less than $0.02\%$. This is well within the theoretical
uncertainty estimated in the last term of eq.~(\ref{interp}), which
implies a shift in $\mh$ of about $150$~MeV. Such a good numerical
agreement is particularly remarkable in view of the fact that the
calculations of refs.~\cite{Buttazzo:2013uya} and
\cite{Kniehl:2015nwa} differ substantially in what concerns the
renormalization of the Higgs vev and the corresponding treatment of
the tadpole contributions (they also differ in the treatment of
higher-order QCD corrections to the top Yukawa coupling). When we move
away from the observed value of the Higgs mass, the discrepancy
between the two calculations varies, reaching up to about $120$~MeV
for the lowest considered value $\mh=120$~GeV. While such discrepancy
remains within the theoretical uncertainty of eq.~(\ref{interp}), the
behavior of the blue line in figure~\ref{fig:buttvsmr} suggests that,
for values of $\mh$ a few GeV away from the observed one, the linear
interpolation loses accuracy and the full dependence on the value of
the Higgs mass should be taken into account.

Finally, we performed an analogous test on the $\mt$ dependence of
eq.~(\ref{interp}), keeping the value of the pole Higgs mass that we
feed into {\tt mr} fixed to $125.15$~GeV, and varying the pole top
mass by $\pm 2$~GeV around its central value of $173.34$~GeV. We find
that the difference between the initial value of $\mh$ and the one
obtained by inserting in eq.~(\ref{interp}) the value of
$\lambda(\mt)$ computed by {\tt mr} varies only by about $10$~MeV in
the considered range of $\mt$. This suggests that the linear
interpolation of the dependence on the pole top mass in
eq.~(\ref{interp}) is not problematic.

\vfill

\section{Impact of the mixed \qcdew\  corrections}
\label{sec:tlnumbers}

In this section we investigate the numerical impact of the mixed
\qcdew\ threshold corrections to the quartic Higgs coupling on the
prediction for the Higgs mass in the MSSM with heavy superpartners. We
use the code {\tt mr}~\cite{Kniehl:2016enc} to extract -- at full
two-loop accuracy -- the $\msbar$-renormalized parameters of the SM
Lagrangian from a set of physical observables, and to evolve them up
to the SUSY scale using the three-loop RGEs of the SM. As mentioned in
the previous section, in our EFT approach the fact that we combine a
full two-loop calculation at the EW scale with an incomplete two-loop
calculation at the SUSY scale does not entail the risk of spoiling
crucial cancellations between large corrections. Throughout the
section we use the world average $\mt =
173.34$~GeV~\cite{ATLAS:2014wva} for the pole top mass, and fix the
remaining physical inputs (other than the Higgs mass) to their current
PDG values~\cite{Tanabashi:2018oca}, namely $G_F= 1.1663787 \times
10^{-5}$~GeV$^{-2}$, $\MZ = 91.1876$~GeV, $\MW = 80.385$~GeV, $\mb =
4.78$~GeV and $\alpha_s(\MZ)=0.1181$.


In order to obtain a prediction for the Higgs mass from a full set of
MSSM parameters, we vary the value of the pole mass $\mh$ that we give
as input to {\tt mr} until the value of the $\msbar$-renormalized SM
parameter $\lambda(Q)$ returned by the code at the SUSY scale $Q=\MS$
coincides with the MSSM prediction of eq.~(\ref{looplam}). In addition
to the mixed \qcdew\ corrections to the quartic Higgs coupling
computed in this paper, we use the results of
refs.~\cite{Bagnaschi:2014rsa, Bagnaschi:2017xid} for the full
one-loop correction $\Delta \lambda^{1\ell}$ and for the ``gaugeless''
part of the two-loop correction $\Delta \lambda^{2\ell}$. We recall
that, as discussed in ref.~\cite{Bagnaschi:2017xid}, we must also
convert the $\msbar$-renormalized bottom Yukawa coupling $g_b(\MS)$
returned by {\tt mr} into its $\drbar$-renormalized MSSM couterpart,
$\hat g_b(\MS)$, to avoid the occurrence of potentially large
$\tb$-enhanced terms in $\Delta \lambda^{2\ell}$. To this effect, we
make use of eqs.~(8), (9), (12) and (13) of
ref.~\cite{Bagnaschi:2017xid}.

\begin{figure}[t]
\begin{center}
\vspace*{-1cm}
\includegraphics[width=13cm]{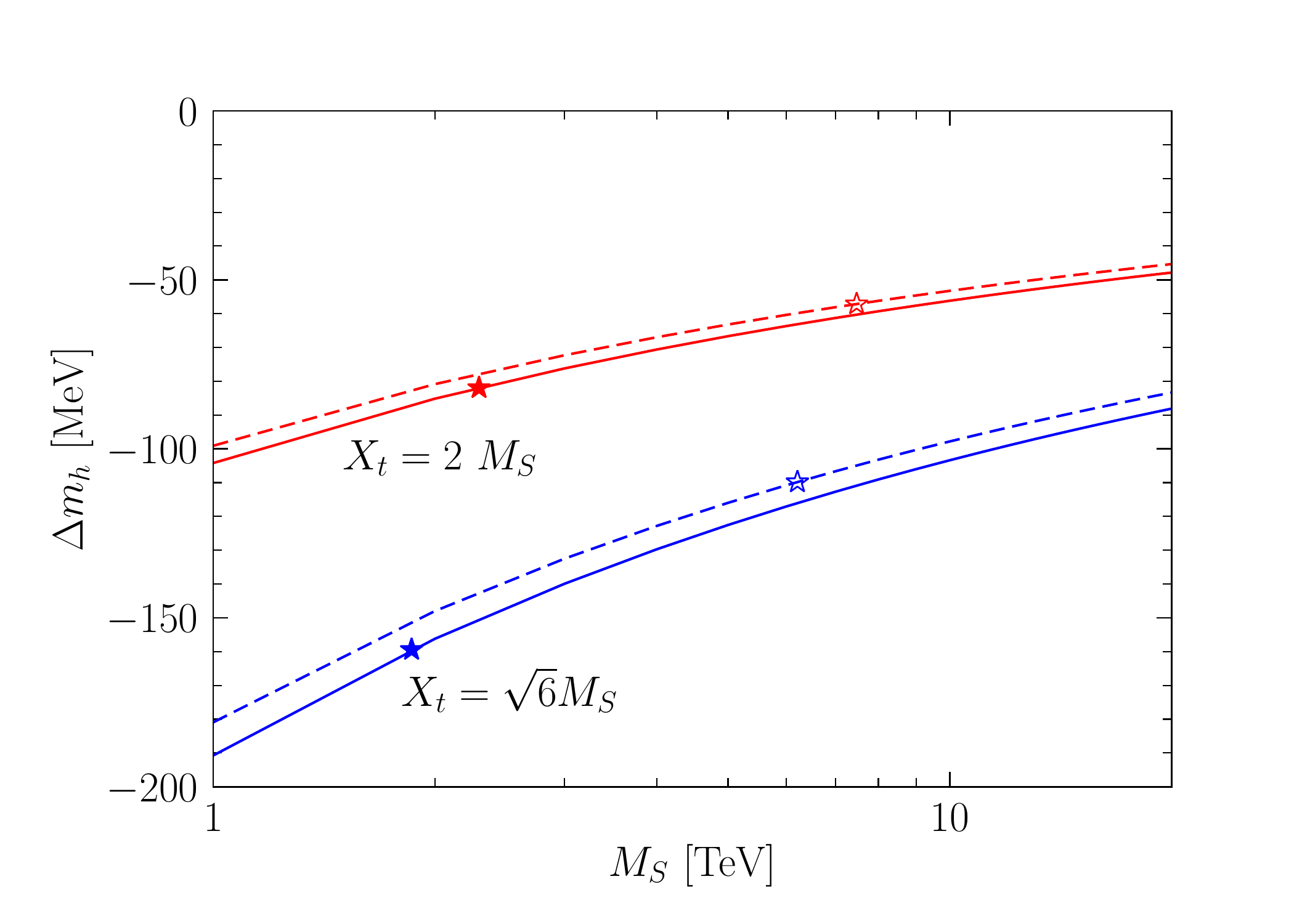}
  \caption{\em Difference (in MeV) between the predictions for the
    Higgs mass obtained with and without the inclusion of the mixed
    \qcdew\  corrections to the quartic Higgs coupling, as a function of
    a common SUSY scale $\MS$, for $X_t=\sqrt{6}\,\MS$ (lower, blue
    lines) or $X_t=2\,\MS$ (upper, red lines), and
    $A_b=A_\tau=A_t$. In each set of lines the solid one is obtained
    with $\tan\beta=20$ and the dashed one with $\tan\beta=5$. The
    star on each line marks the value of $\MS$ for which the improved
    calculation of $\Delta \lambda^{2\ell}$ leads to $\mh =
    125.09$~GeV.}
  \label{fig:dmh}
\vspace*{-3mm}
\end{center}
\end{figure}

In figure~\ref{fig:dmh} we show the difference (in MeV) between the
EFT predictions for the Higgs mass obtained with and without the
inclusion of the mixed \qcdew\ corrections to the quartic Higgs
coupling. We consider a simplified MSSM scenario in which the masses
of all superparticles (sfermions, gauginos and higgsinos) as well as
the mass of the heavy Higgs doublet are set equal to the common SUSY
scale $\MS$, which we vary between $1$~TeV and $20$~TeV. The
left--right stop mixing parameter is fixed either as $X_t=\sqrt 6
\,\MS$ (lower, blue lines) or $X_t=2\,\MS$ (upper, red lines). In each
set of lines the solid one is obtained with $\tan\beta=20$ and the
dashed one with $\tan\beta=5$.  The remaining input parameters are the
trilinear Higgs--sfermion interaction terms for sbottoms and staus,
which we fix as $A_b=A_\tau=A_t$. We remark that all of the MSSM
parameters are interpreted as $\drbar$-renormalized quantities
expressed at the renormalization scale $Q=\MS$. The star on each line
marks the value of $\MS$ for which the improved calculation of $\Delta
\lambda^{2\ell}$ (i.e., including the mixed \qcdew\ corrections) leads
to the observed value of the Higgs mass, $\mh =
125.09$~GeV~\cite{Aad:2015zhl}.

Figure~\ref{fig:dmh} shows that, in the considered scenario, the
mixed \qcdew\  corrections computed in this paper are fairly small,
shifting the MSSM prediction for $\mh$ downwards by ${\cal
  O}(100)$~MeV. The fact that the corrections are reduced (in absolute
value) for larger values of $\MS$ is partially due to the scale
dependence of the relevant couplings: with the exception of $\gp$,
they all decrease with increasing $Q=\MS$. The comparison between the
solid and dashed lines in each set shows that the dependence of the
mixed \qcdew\  corrections on $\tb$ is rather mild (in contrast, the
overall prediction for $\mh$ depends strongly on $\tb$, as shown by
the relative position of the stars on the solid and dashed
lines). Finally, the comparison between the lower (blue) and upper
(red) sets of lines shows that the mixed \qcdew\  corrections depend
rather strongly on the ratio $|X_t/\MS|$, with smaller ratios leading
to smaller corrections. Indeed, we checked that for $|X_t/\MS| < 1$
the effect of the corrections can be at most of ${\cal O}(10)$~MeV in
the considered scenario.

\begin{figure}[t]
\begin{center}
\vspace*{-1cm}
\includegraphics[width=12.5cm]{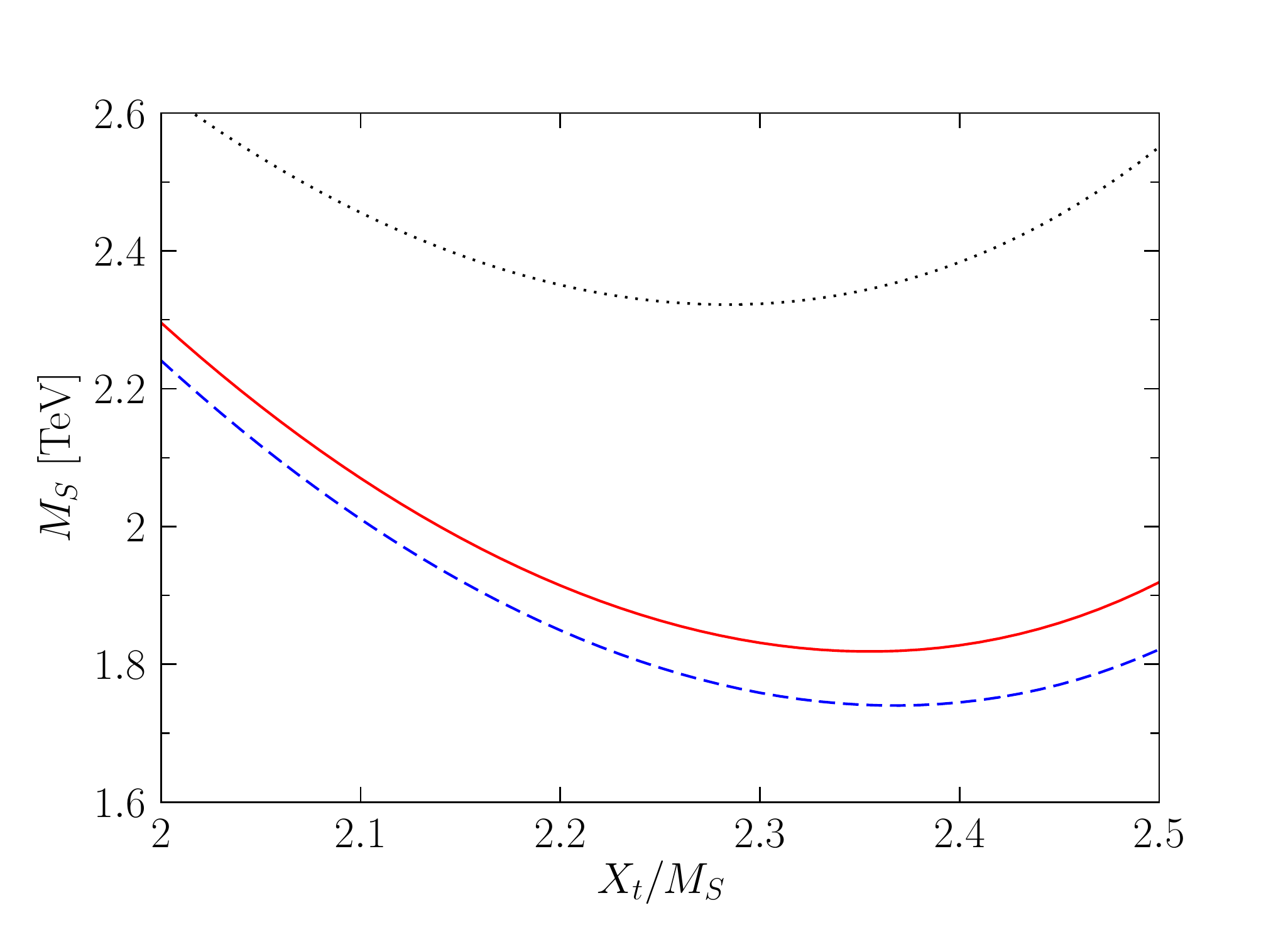}
  \caption{\em Values of the common SUSY scale $\MS$ and of the stop
    mixing term $X_t$ that lead to $\mh = 125.09$~GeV, for
    $\tan\beta=20$ and $A_b=A_\tau=A_t$. The dotted line is obtained
    including only the one-loop threshold corrections to the Higgs
    quartic coupling, the dashed line includes the two-loop
    corrections in the gaugeless limit, and the solid line includes
    also the effect of the mixed \qcdew\  corrections.}
  \label{fig:MSXt}
\vspace*{-1mm}
\end{center}
\end{figure}

An alternative way to assess the effect of the newly-computed
corrections consists in taking the measured value of the Higgs mass as
an input parameter, and using the matching condition on the quartic
Higgs coupling at the SUSY scale, eq.~(\ref{looplam}), to constrain
the MSSM parameters. In figure~\ref{fig:MSXt} we show the values of
$\MS$ and $X_t$ that lead to $\mh = 125.09$~GeV, in the simplified
scenario with degenerate superparticle and heavy-Higgs masses, for
$\tan\beta=20$ and $A_b=A_\tau=A_t$. We focus on values of the ratio
$X_t/\MS$ between $2$ and $2.5$, which allow for SUSY masses around 2
TeV (i.e., roughly at the limit of the HL-LHC
reach~\cite{CidVidal:2018eel}). Once again, all of the MSSM parameters
are interpreted as $\drbar$-renormalized quantities expressed at the
renormalization scale $Q=\MS$.  The (black) dotted line in
figure~\ref{fig:MSXt} is obtained including only the one-loop
threshold corrections to the Higgs quartic coupling, the (blue) dashed
line includes the two-loop corrections in the gaugeless limit, and the
(red) solid line includes also the effect of the mixed \qcdew\ 
corrections. Unsurprisingly, the comparison between the three lines
shows that the mixed \qcdew\  corrections are sub-dominant with respect
to the two-loop corrections computed in the gaugeless
limit. Nevertheless, they can shift the value of $\MS$ that leads to
the observed Higgs mass by ${\cal O}(100)$~GeV in the direction of
heavier superparticles.
    
\section{Conclusions}
\label{sec:conclusions}

If the MSSM is realized in nature, both the measured value of the
Higgs mass and the negative results of the searches for superparticles
at the LHC suggest some degree of separation between the SUSY scale
and the EW scale. In this scenario, the MSSM prediction for the Higgs
mass is subject to potentially large logarithmic corrections, making a
fixed-order calculation of $\mh$ inadequate and calling for an
all-orders resummation in the EFT approach.

In this paper we improved the EFT calculation of the Higgs mass in the
MSSM, by computing the class of two-loop threshold corrections to the
quartic Higgs coupling that involve both the strong and the EW gauge
couplings. Combined with the ${\cal O}(g_t^4\,g_s^2)$ and ${\cal
  O}(g_b^4\,g_s^2)$ corrections previously provided in
refs.~\cite{Bagnaschi:2014rsa, Bagnaschi:2017xid}, this completes the
calculation of the two-loop threshold corrections that involve the
strong gauge coupling.  Our calculation involves novel complications
with respect to the case of the ``gaugeless'' two-loop corrections of
refs.~\cite{Bagnaschi:2014rsa, Bagnaschi:2017xid}. While in the latter
all two-loop diagrams could be computed in the effective potential
approach (i.e., for vanishing external momenta), the mixed \qcdew\ 
corrections include contributions from the ${\cal O}(p^2)$ parts of
the two-loop self-energies of the Higgs and gauge bosons. We obtained
results for the threshold corrections to the quartic Higgs coupling
valid for generic values of all the relevant SUSY parameters, which we
make available on request in electronic form. For the sake of
illustration, in section~\ref{sec:matching} we provided explicit
formulae in the simplified limit of degenerate superparticle masses.

We remark that our calculation can be trivially adapted also to the
split-SUSY scenario in which the gluino is much lighter than the
squarks, by taking the limit of vanishing gluino mass in our full
results. On the other hand, in scenarios in which the gluino is
heavier than the squarks the two-loop corrections to the quartic Higgs
coupling contain potentially large terms enhanced by powers of the
ratios between the gluino mass and the squark masses. This is a
well-known aspect of the $\drbar$ renormalization of the squark masses
and trilinear couplings~\cite{Degrassi:2001yf, Vega:2015fna,
  Bagnaschi:2017xid, Braathen:2016mmb}, which could be addressed
either by devising an ``on-shell'' scheme adapted to the heavy-SUSY
setup, or by building a tower of EFTs in which the gluino is
independently decoupled at a higher scale than the squarks.

In the EFT approach, the inclusion of new threshold corrections to the
quartic Higgs coupling at the SUSY scale mandates that corrections of
the same perturbative order in the relevant couplings be included in
the calculation of the pole Higgs mass at the EW scale. In the
simplest heavy-SUSY setup in which the effective theory valid below
the SUSY scale is just the SM, we can exploit the full two-loop
calculations of the relation between $\mh$, $\lambda(\qew)$ and $G_F$
presented in refs.~\cite{Buttazzo:2013uya, Kniehl:2015nwa,
  Martin:2019lqd}. In section~\ref{sec:weakscale} we compared the
results of two of those calculations, refs.~\cite{Buttazzo:2013uya}
and \cite{Kniehl:2015nwa}, discussing the range of validity of an
interpolating formula provided in ref.~\cite{Buttazzo:2013uya}.

In section~\ref{sec:tlnumbers} we investigated the numerical impact of
the mixed \qcdew\  corrections to the quartic Higgs coupling.  We
considered a simplified MSSM scenario with degenerate masses for all
superparticles and for the heavy Higgs doublet, focusing on the region
of the parameter space in which the prediction for the Higgs mass is
close to the observed value and the stop squarks are in principle
still accessible at the HL-LHC. We used the code {\tt
  mr}~\cite{Kniehl:2016enc}, based on the calculation of
ref.~\cite{Kniehl:2015nwa}, to extract all of the SM couplings from a
set of physical observables and to evolve them up to the SUSY scale,
where we compare the value of the quartic Higgs coupling with its MSSM
prediction. We found that the impact of the newly-computed two-loop
corrections on the prediction for the Higgs mass tends to be small,
and it is certainly sub-dominant with respect to the impact of the
``gaugeless'' two-loop corrections. In the considered scenario, the
mixed \qcdew\  corrections can shift the prediction for the Higgs mass
by ${\cal O}(100)$~MeV, and they can shift the values of the stop
masses required to obtain the observed value of $\mh$ by ${\cal
  O}(100)$~GeV.

We stress that the smallness of these effects is in fact a desirable
feature of the EFT approach to the calculation of the Higgs mass.
While the logarithmically enhanced corrections are accounted for by
the evolution of the parameters between the matching scale and the EW
scale, and high-precision calculations at the EW scale can be borrowed
from the SM, the small impact of new two-loop corrections computed at
the SUSY scale suggests that the uncertainty associated to uncomputed
higher-order terms should be well under control in the considered
scenario.
On the other hand, we recall that the accuracy of the measurement of
the Higgs mass at the LHC has already reached the level of
$100\!-\!200$~MeV~\cite{Tanabashi:2018oca} -- i.e., it is comparable
to the effects of the corrections discussed in this paper -- and will
improve further when more data are analyzed. If SUSY eventually shows
up at the TeV scale, the mass and couplings of the SM-like Higgs boson
will serve as precision observables to constrain MSSM parameters that
might not be directly accessible by experiment. To this purpose, the
accuracy of the theoretical predictions will have to match the
experimental one, making a full inclusion of two-loop effects in the
Higgs-mass calculation unavoidable. Our results should be viewed as a
necessary step in that direction.

\section*{Acknowledgments}

We thank Mark Goodsell for useful discussions.  The work of S.~P.~and
P.~S.~is supported in part by French state funds managed by the Agence
Nationale de la Recherche (ANR), in the context of the LABEX ILP
(ANR-11-IDEX-0004-02, ANR-10-LABX-63) and of the grant
``HiggsAutomator'' (ANR-15-CE31-0002). G.~D.~acknowledges warm
hospitality at LPTHE and support from ``HiggsAutomator'' during the
completion of this work.

\vfill
\newpage


\bibliographystyle{utphys}
\bibliography{BDPS}

\end{document}